\documentclass[useAMS,usenatbib]{mnras}
\usepackage{longtable}
\usepackage{graphicx}
\usepackage{lscape}
\usepackage{rotating}

\newcommand{\gppr}{\stackrel{>}{\scriptstyle \sim}}
\newcommand{\gappr}{\raisebox{-0.4ex}{$\gppr$}}
\newcommand{\lppr}{\stackrel{<}{\scriptstyle \sim}}
\newcommand{\lappr}{\raisebox{-0.4ex}{$\lppr$}}

\title[Axions and the luminosity function of white dwarfs] {Axions and
  the luminosity function  of white dwarfs. The thin  and thick disks,
  and the halo}

\author[J. Isern  et  al.]{J. Isern$^{1,2}$,
  E. Garc\'ia-Berro$^{3,2,\dagger}$,
  S.  Torres$^{3,2}$,
  R. Cojocaru$^{3,2}$,
  S. Catal\'an$^{4,2}$\\
$^{1}$ Institute  of Space Sciences  (ICE, CSIC), Campus UAB,  Cam\'{\i} de
  Can  Magrans  s/n,  08193  Cerdanyola  del  Vall{\`e}s  (Barcelona),
  Spain\\
  $^{2}$  Institute for  Space Studies  of Catalonia,  c/Gran Capit\`a
  2--4, Edif. Nexus 201, 08034 Barcelona, Spain\\
$^{3}$  Departament  de  F\'\i   sica,  Universitat  Polit\`ecnica  de
  Catalunya, c/Esteve Terrades 5, 08860 Castelldefels, Spain\\
$^{4}$ Center for Astronomy, National University of Ireland, Newcastle Road,
  SW4 794 Galway, Ireland\\
$^{\dagger}$ Deceased September 23rd, 2017}

\begin{document}
\date{Accepted .... Received 2017; in original form 2017}
\pagerange{\pageref{firstpage}--\pageref{lastpage}} \pubyear{2017}
\maketitle

\begin{abstract}
  The evolution  of white dwarfs  is a simple gravothermal  process of
  cooling.  Since the shape of  their luminosity function is sensitive
  to the characteristic cooling time, it  is possible to use its slope
  to test the existence of additional sources or sinks of energy, such
  as those predicted by alternative physical theories. The aim of this
  paper is  to study if  the changes in the  slope of the  white dwarf
  luminosity function  around bolometric magnitudes ranging  from 8 to
  10 and previously  attributed to axion emission  are, effectively, a
  consequence  of  the  existence  of   axions  and  not  an  artifact
  introduced  by  the star  formation  rate.   We compute  theoretical
  luminosity functions of the thin and  thick disk, and of the stellar
  halo including axion emission and  we compare them with the existing
  observed luminosity functions.  Since these stellar populations have
  different star  formation histories, the slope change  should be
  present in all of  them at the same place if it is  due to axions or
  any other intrinsic cooling mechanism. The signature of an unexpected
  cooling seems to be  present in the luminosity functions  of the thin
  and thick  disks, as well as  in the halo luminosity  function. This
  additional  cooling   is  compatible   with  axion   emission,  thus
  supporting to the idea that DFSZ axions, with a mass in the range of
  4 to  10~meV, could exist. If  this were the  case, these axions  could be
  detected by the future solar axioscope IAXO.
\end{abstract}


\label{firstpage}


\section{Introduction}

White   dwarfs  are   the  final   evolutionary  stage   of  low   and
intermediate-mass  stars  -- namely,  those  with  $M\leq 10\pm  2  \,
M_{\sun}$   \citep{alth10}.    They   are  degenerate   objects   and,
consequently, they  cannot obtain energy from  thermonuclear reactions
and their evolution is just  a gravothermal process of cooling.  Their
structure  is relatively  simple.  They  have a  degenerate core  that
contains the  bulk of mass  and acts as a  reservoir of energy,  and a
partially degenerate envelope that controls the energy outflow. White dwarfs 
 with masses $M\la 0.4  \, M_{\sun}$ have He  cores, while
those with  a $M \ga 1.1  \, M_{\sun}$ have O-Ne  cores. Finally, those
with intermediate  masses, which represent the vast  majority, have  a core
made of a mixture of C and O plus some impurities like $^{22}$Ne.  All
of them are surrounded by a thin  He layer with a mass $\sim 10^{-2}\,
M_{\rm WD}$.   However, in $\sim 75$\%  of the cases, these  stars are
also  surrounded by  an even  thinner layer  of hydrogen  with a  mass
between $10^{-4}$ and $10^{-15}\,  M_{\sun}$.  White dwarfs displaying
hydrogen in their spectra are known  as DA and the remaining ones are
generically referred to as non--DA.

Simple theoretical  considerations show that the  coolest white dwarfs
have ages  $\gappr \,10$~Gyr.  Therefore,  they allow to  obtain important
information about the past history  of the Galaxy, in particular about
the age of the disk, the halo and the system of globular and open clusters, as
well as the  star formation  rate -- see the reviews by 
\citet{koes90,iser98a,font01,koes02,hans03,hans04,alth10,iser13,garc16}.   
Moreover, it is  important to realize that the accuracy and  reliability  
of  the  formation  history  of  all  these  stellar components will be 
much improved when the data obtained by {\sl Gaia} will become available \citep{bars14}.

\begin{figure}
\centering
\includegraphics[width=\hsize]{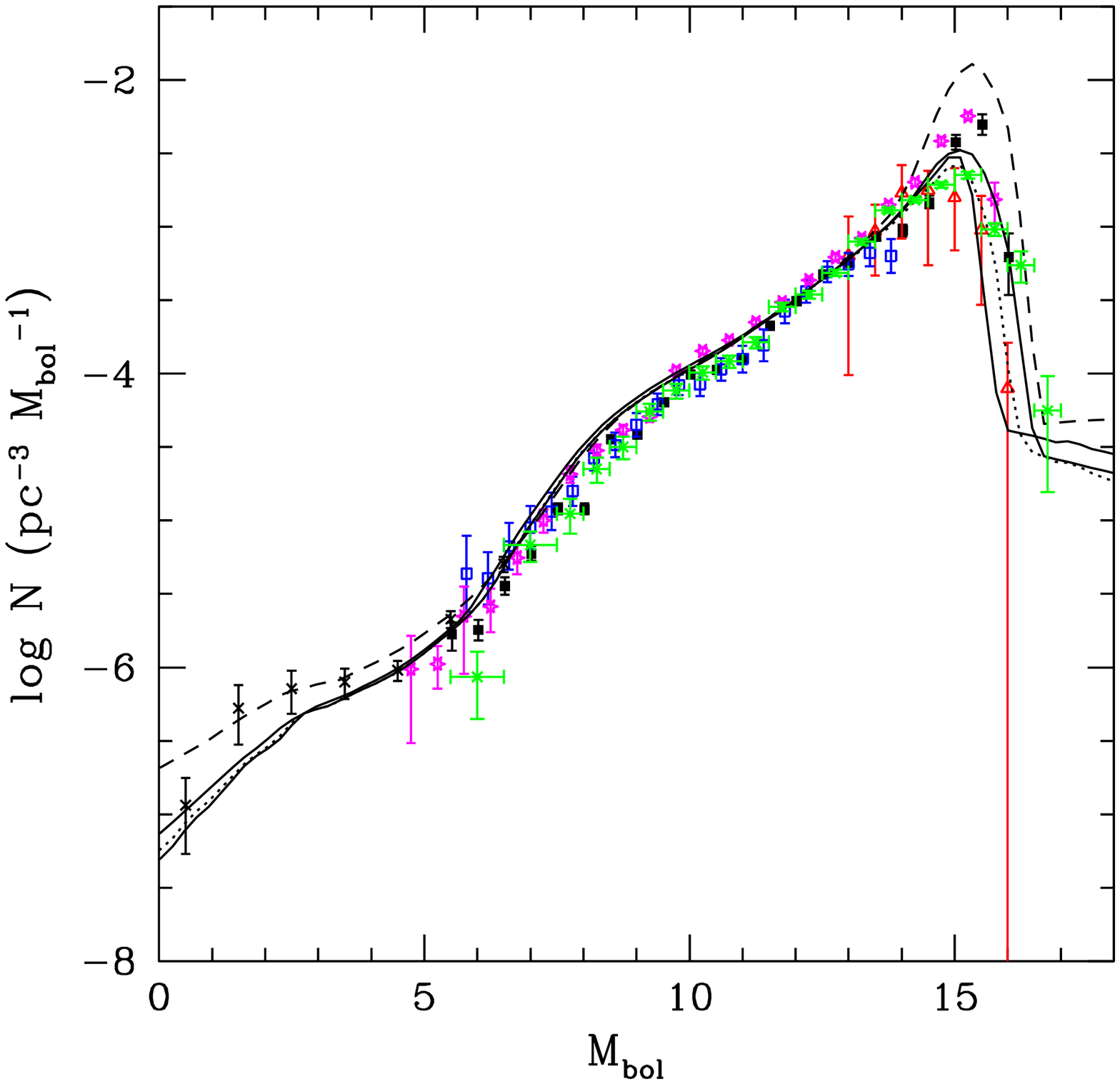}
\caption{ Luminosity  functions  obtained  from different authors:
black solid squares \citep{harr06},  open blue squares \citep{dege08}, black crosses \citep{krze09}, magenta stars \citep{rowe11} and green stars \citep{munn17}.  The two solid black  lines represent the  theoretical  luminosity functions  obtained   assuming  a  constant SFR and an age of  the Galaxy of 10 and 13~Gyr, respectively. The dashed line represents the one obtained with  a decreasing exponential SFR and, finally, the dotted line is obtained using an almost constant SFR with an exponential tail. See text for further details. 
}
\label{fig1}
\end{figure}

The tool to  obtain such information is the  luminosity function, i.e.
the distribution  of white dwarfs  as a function of  their luminosity (see Fig. \ref{fig1}). 
This function was derived for the first time by \citet{weid68}, about four decades ago, and 
was improved by several authors during the nineties \citep{lieb88,evan92,oswa96,legg98,knox99}. 
These early luminosity functions proved that the evolution of white dwarfs is just a cooling 
process and that there was a cut-off in the distribution caused by the finite age of the Galaxy.

The position of the cut-off is sensitive to the cooling rate and, consequently, it 
can be used to constrain any new theory  or hypothesis implying the introduction of an additional 
source or sink of energy. However the low number of stars in the samples, few hundreds, and 
the uncertainties in the position of the cut-off prevented anything else than obtaining upper 
bounds. This was the cases of the mass of the axion \citep{raff96} or the secular drift of the 
Newton constant \citep{garc95}, just to cite two examples. 

The advent of large cosmological surveys like the Sloan Digital Sky Survey (SDSS) 
introduced a noticeable improvement in the precision and accuracy of the luminosity function of 
white dwarfs since they allowed to increase the sample size to several thousands of stars. Fig.~\ref{fig1} shows the luminosity function of disk white dwarfs obtained by several authors. 
Black solid squares represent the data corresponding to $\sim 6000$ DA and non-DA white dwarfs obtained from the SDSS Data Release 3 and the USNO-B catalogues \citep{harr06} using the proper motion method. A weak point of the luminosity functions obtained from the SDSS catalogues is that the integration time is fixed and dim white dwarfs systematically have poorer S/N ratios \citep{limo10}. Fortunately, the Super COSMOS Sky Survey, a completely independent survey that also uses proper motion to identify white dwarfs, is completely free of this systematic effect. The luminosity function obtained from this catalogue is represented by magenta stars \citep{rowe11} in Fig.~\ref{fig1}. As can be seen, both functions nearly 
coincide over a large part of the brightness interval 
when normalized to $\log (L/L_\odot) \approx  -3$ or $M_{\rm bol} \approx 12$. At large brightness, $M_{\rm bol} \lappr 6$, both 
luminosity functions display a large dispersion, not plotted in the figure, as a consequence of 
the fact that the proper motion method is not appropriate there. One way to circumvent 
this problem is using the UV-excess technique \citep{krze09}. The results obtained with this 
technique are represented by black crosses in Fig.~\ref{fig1}  after matching their dim region 
with the corresponding bright segment  of the \citet{harr06} distribution. As 
a complement, the luminosity function of the dimmest white dwarfs obtained by \citet{legg98} 
has also been included in Fig.~\ref{fig1} as red triangles.  The discrepancies at low luminosities are due to the difficulty to separate DAs from non-DAs and to the different behavior of the envelope. 
The green crosses represent the luminosity function recently obtained by \citet{munn17}, also from the SDSS, that represents a noticeable improvement of the data obtained by \citet{harr06}. Finally, the blue hollow squares represent the luminosity function of only DA stars \citep{dege08} normalized to the other luminosity functions at $\log (L/L_\odot) = -3$ in order to compare the slopes.

The improved quality of the new luminosity functions has allowed, for the first time, to 
determine their shape and to use the slope as a tool to test new physical theories  \citep{raff96,iser08}.  If  an additional  source or  sink  of energy  is  added to the energy budget of the white dwarf, the  characteristic cooling time  is modified  and its imprint  appears in  the luminosity function, as can  be seen  in Fig.~\ref{fig1},  where the  change of slope that appears at magnitudes $\sim 8$ is caused by the transition from the cooling dominated by neutrinos to the one dominated by photons. As an example, this technique was used by \citet{iser08a} to suggest that axions of the DFSZ type could be contributing to the cooling of white dwarfs.

One  of the  main uncertainties  comes  from the  star formation  rate
(SFR).  If  this function  is relatively smooth  and does not increase
abruptly with time,  the shape of the bright branch  of the luminosity
function is  entirely determined by  the physical properties  of white
dwarf  stars \citep{iser08}.   It has  been shown  \citep{rowe13} that, by
adjusting  bin  by bin  it  is  possible to  find  a  SFR that  almost
completely  fits the  observed WDLF  and is compatible with  SFRs 
obtained with independent methods \citep{cign06}.

This  SFR, however,  has been  obtained as  an ``ad  hoc'' fit  to the
observations, employing the existing white dwarf models. Consequently,
other  models  with  different   physical  ingredients  would  provide
different solutions. One way to break this degeneracy and to decide if
it is  necessary to include new  physics is to examine  the luminosity
functions  of  populations  that   have  different  stellar  formation
histories. In this case, if  axions (or any other physical effect) 
are responsible for this anomalous cooling,  their  signature would  be  present  in all  the  luminosity
functions  at  roughly  the  same  luminosities.  

Another important uncertainty comes from the adopted white dwarf scale 
height at the moment of estimating their density. \citet{harr06} showed that 
changing the scale height has a noticeable effect on the shape of the bright 
branch of their luminosity function.  Since the \citet{munn17} luminosity function is 
representative of a mixture of thin and thick disk stars, \citet{kili17} have also shown that it is 
possible to remove the mentioned discrepancy between the observational and 
theoretical luminosity functions by just adopting a variable scale height 
connecting the thin and thick disk populations. This argument, that 
is correct, reinforces the convenience of analyzing the luminosity function of 
white dwarfs belonging to populations with independent scale heights and with different star 
formation histories to disentangle the intrinsic and group properties.

One possibility is provided by the ordinary galactic populations. The common picture is
that spiral galaxies form inside a spinning dark matter halo. In the first stage, the gas collapses 
in a dynamical time scale of $\sim 10^8$~yrs, leaving behind a stellar halo,
and settles down into a disk from where the stars form. Galactic disks can be easily destroyed
by mergers with other structures of similar mass. Therefore, the existence of an undamaged disk 
implies that it has no suffered important mergers since it was born. Furthermore, if an early 
formed disk is perturbed by a minor merger, it can form a thick disk which, in turn, can lead 
to the formation of a thin disk inside it. According to this picture, the present structure of 
the Milky Way could be the result of the following sequence of events \citep{reid05}: i) formation 
of a primitive halo at $t \sim 11-12$ Gyr; ii) episodes of minor mergers from the satellite galactic system; iii) formation of the disk at $t \sim 10 - 11$ Gyr; iv) formation of a thick disc at $ t \sim 9-10$ Gyr; v) formation of the thin disk at $ t \sim 8$ Gyr. Therefore, it seems reasonable to assume that the SFR of the inner halo, and of the thin and thick disks are reasonably independent.  Using kinematic  arguments  it  is  possible to  distinguish  between the white dwarfs that belong  to  these three structures  \citep{rowe11,munn17} and, consequently it is natural to look for any anomalous cooling in the  region $M_{\rm bol} \sim 7-12$ of the luminosity functions of these white dwarf populations. In particular, the axion case will be analysed as an example.

Another possibility is provided by globular clusters \citep{hans15,garc14,gold16}, but the analysis is not straightforward since globulars are not made of a single stellar population but of two or even three \citep{grat12}. Furthermore, the dynamics of the cluster, sink of heavy stars towards the center for instance, can be strongly modified by the presence of moderately massive black holes in the center, as is the case of 47 Tuc \citep{kizi17}. For these reasons, we will analyze the case of globular clusters in a separate paper.


\section{The luminosity function of white dwarfs}

\begin{figure}
\centering
\includegraphics[width=\hsize]{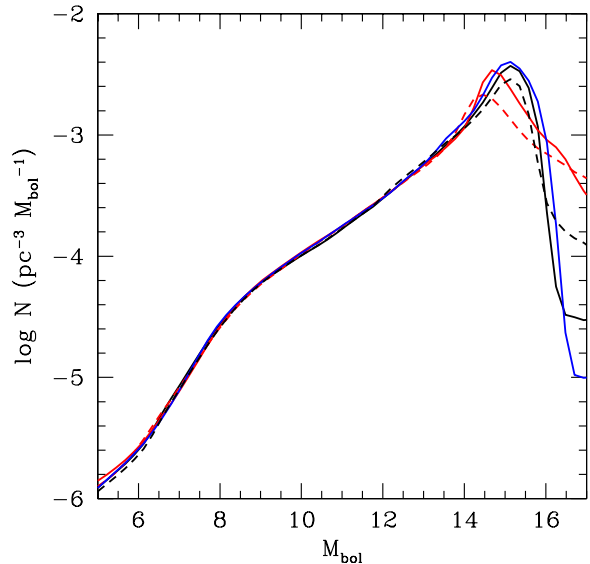}
\caption{Theoretical luminosity functions obtained with the C--O white  dwarf models 
of:  \citet{sala10} -- continuous, pure DA stars, and dashed, 80\% DAs and 20\%non-DAs,   black lines,  \citet{rene10} -- blue line -- and \citet{font01} -- continuous, thick H-envelope, and dashed, thin H-envelope, red lines.}
\label{fig2}
\end{figure}

The white dwarf  luminosity function is defined as  the number density
of white dwarfs of a given magnitude per unit magnitude interval:
\begin{equation}
n(l) = \int^{M_{\rm u}}_{M_{\rm l}}\,\Phi(M)\,\Psi(t)
\tau_{\rm cool}(l,M) \;dM
\label{lf}
\end{equation}
In  this  expression,  $t$  satisfies   the  condition  $t  =  T-t_{\rm
cool}(l,M)-t_{\rm PS}(M)$,  where $l  = -\log (L/L_\odot)$,  $M$ is
the mass  of the  parent star  (for convenience  all white  dwarfs are
labeled with the mass of the main sequence progenitor), $t_{\rm cool}$
is   the   cooling   time   down   to   luminosity   $l$,   $\tau_{\rm
cool}=dt/dM_{\rm  bol}$ is  the characteristic  cooling time,  $t_{\rm
PS}$ is the lifetime of the progenitor  of the white dwarf, and $T$ is
the age  of the population under  study.  $M_{\rm u}$ and  $M_{\rm l}$
are the maximum and the minimum masses of the main sequence stars able
to produce  a white  dwarf of luminosity  $l$. Therefore,  $M_{\rm l}$
satisfies the condition
\begin{equation} 
T = t_{\rm cool}(l,M_{\rm l}) + t_{\rm PS}(M_{\rm l})
\label{tml}
\end{equation}
The remaining  quantities, the  initial mass function,  $\Phi(M)$, and
the star formation rate, $\Psi(t)$, are  not known a priori and depend
on the  properties of the  stellar population under study.   Since the
total density of  white dwarfs is not yet well  determined, to compare
theory and observations  it is customary to normalize  the white dwarf
luminosity function to the bin with the smallest error bar, usually at
$l \simeq 3$.  Finally, although it  does not appear explicitly in the
previous  expressions,  it  is  necessary  to  provide  a  relationship
relating  the  mass   of  the  white  dwarf  with  the   mass  of  its
progenitor. In summary, if the stellar properties are well known it is
possible to  obtain information about the  galactic properties, specifically
the star formation rate and the age of the Galaxy. Yet this is not the
only  possibility,  since  the  bright   branch  of  the  white  dwarf
luminosity function -- that with bolometric magnitude $M_{\rm bol} \la
13$ -- is almost independent of the assumed star formation rate.  This
can be explained employing simple arguments.  Eq.~(\ref{lf}) can be written as:

\begin{equation}
n(l) = \left\langle\tau_{\rm cool}\right\rangle 
\int_{M_{\rm l}}^{M_{\rm u}}\phi(M)
\psi(T-t_{\rm cool}-t_{\rm PS})\;dM.
\label{lf1} 
\end{equation}

Restricting  ourselves to  bright white  dwarfs --  namely, those  for
which $t_{\rm  cool}$ is small --  the lower limit of  the integral is
satisfied  by low-mass  stars  and,  as a  consequence  of the  strong
dependence of the main sequence lifetimes with mass, it adopts a value
that  is almost  independent  of the  luminosity under  consideration.
Therefore,  if $\psi$  is a  well-behaved  function and  $T$ is  large
enough,  the lower limit is almost independent of the luminosity, 
and the value of the integral is incorporated  into  the normalization  
constant  in such a way that  the shape  of  the
luminosity function  only depends on the  averaged physical properties
of the white dwarfs \citep{iser08,iser09}.  This average is dominated by low mass white dwarfs and, as far as the mass spectrum is not strongly perturbed by the adopted star formation or the initial mass function, it is representative of the intrinsic properties of white dwarfs. 

It is important to emphasize here that the luminosity band we are considering,
 $6\, \lappr \, M_{\rm bol} \, \lappr\, 13$ is the one with less theoretical 
uncertainties and where all the present models almost coincide \citep{hans03}. In 
particular, Fig.~\ref{fig2} shows,  after normalization, the different luminosity functions obtained with the BaSTI models\footnote{C/O abundances stratified, $M_{\rm He}=10^{-2}M_{\rm wd}$, $M_{\rm H} = 10 ^{-4}M_{\rm wd}$ for DA models, and $M_{\rm He}=10^{-3.5}M_{\rm wd}$. Publicly available at: \url{http://albione.oa-teramo.inaf.it}} \citep{sala10}, 
La Plata models\footnote{C/O abundances stratified,masses of the envelope different for each mass of the white dwarf and in the range $10^{-2.486} \la M_{\rm He} \la 10^{-1.567}M_{\rm wd}$, $10^{-486} \la M_{\rm H} \la  10 ^{-3.441}M_{\rm wd}$ for DA models.
Publicly available at \url{http://www.fcaglp.unlp.edu.ar/~althaus/}} \citep{rene10}, 
and Montreal models\footnote{C/O  half and half and uniformly distributed in the star.
Thick envelopes $M_{\rm He}=10^{-2}M_{\rm wd}$, $M_{\rm H} = 10 ^{-4}M_{\rm wd}$, and thin envelopes $M_{\rm He}=10^{-2}M_{\rm wd}$, $M_{\rm H} = 10 ^{-10}M_{\rm wd}$. Publicly available at \url{http://www.astro.umontreal.ca/~bergeron/CoolingModels}} \citep{font01}.
These luminosity functions were computed  as in \citet{iser98} and in \citet{mill14}, using the initial final mass relationship of \citet{cata08}.

 As can be seen, these luminosity functions are almost indistinguishable in the region $M_{\rm bol} \la 13$. The differences due to the models themselves or the
shape of  the SFR only appear  in the regions containing  cool or very
bright  white   dwarfs.   Unfortunately,  the   present  observational
uncertainties in these regions prevent at present to discriminate among the different possibilities. In the following we will 
adopt the BaSTI models as the reference for calculations. A 
discussion on the sensitivity of the luminosity function to the different inputs 
can be found in \citet{iser09} and in \url{http://astro.uni-tuebingen.de/~rauch/EUROWD10.html} 

In the bright region considered here, the slope of the relationship between the luminosity and the core temperature of DA and non--DA white dwarfs almost coincide, see Fig. 1 of \citet{sala10}. If we adopt a relationship of the form $L = gT_{\rm c}^\gamma$, where  $\gamma = \gamma _{\rm DA} \approx \gamma _{\rm nDA}$ in the luminosity range mentioned before, and we neglect the compression term,
$L \approx  - dU/dt \approx  - C_{\rm V}\,dT_{\rm c}/dt$,
we obtain 
$n(l) \propto {\dot l}^{ - 1} =  - L(dL/dt)^{-1} = C_{\rm V}(g\gamma)^{-1}T_C^{1 - \gamma }$, and the DA and non--DA luminosity functions almost overlap in this luminosity interval after normalization. This is the reason why the luminosity function of \citet{dege08} containing only DAs coincides with those containing DAs and non--DAs after normalization.

Fig.~\ref{fig1} also displays  the theoretical disk white  dwarf luminosity 
functions obtained assuming  a  constant SFR but two ages of  the Galaxy of 10 and 13~Gyr, 
respectively, a decreasing exponential SFR, $\Psi \propto \exp (-t/\tau),\, \tau= 3\,{\rm Gyr} $, 
and, finally, an almost constant SFR with an exponentially decreasing tail that represents models 
where the star formation propagated from the center to the periphery,
 $\Psi \propto (1+\exp [(t-t_0)/\tau])^{-1}, \tau = 3 \, {\rm Gyr},\, t_0= 10\,{\rm Gyr}$. 
As can be seen, in the region $6 \, \lappr\, M_{\rm bol}\, \lappr \, 13$, all luminosity 
functions overlap as far as the SFR is smooth enough.

\begin{figure}
\centering
\includegraphics[width=\hsize]{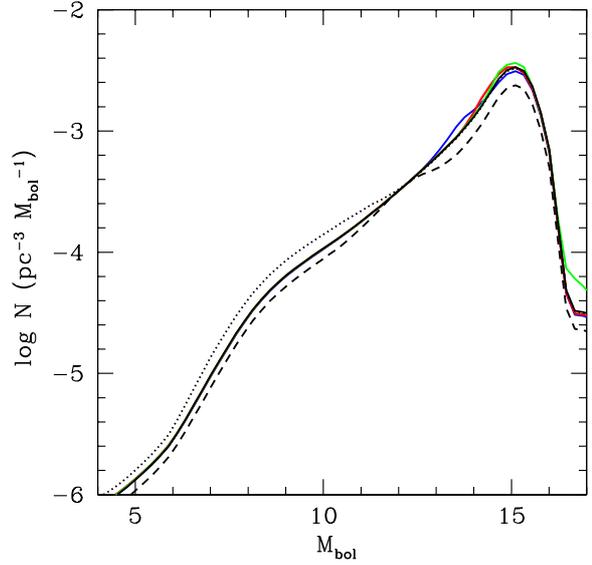}
\caption{Signatures  in the  disk white  dwarf luminosity  function of
  different  bursts  of  star  formation.  The  solid  lines  show  the
  luminosity function for a constant  star formation rate (black line)
  and for  a constant star  formation rate  with three bursts  of star
  formation occurring 4, 6 and 8~Gyr  ago (blue, red and green lines).
  The dotted and dashed lines  correspond to a constant star formation
  rate with bursts occurring 1  and 2~Gyr ago, respectively.  See text
  for additional details.}
\label{fig3}
\end{figure}

Now we go one step forward and we ask ourselves for the effects on the
luminosity function of short bursts  of star formation.  This question
is  important   as  it  may   modify  the  slope  of   the  luminosity
function \citep{noh90}. For  instance, when  a young  burst is  considered, low-mass
main sequence stars have no time to become white dwarfs. Since $M_{\rm
l}$ in  Eq.~(\ref{lf1}) depends  on the luminosity,  the slope  of the
luminosity  function  is  affected.   Fig.~\ref{fig3}  represents  the
luminosity functions obtained when a  burst of star formation is added
to a  constant SFR,  $\psi(t)=\psi_0 [1+\delta(t_1,\Delta  t)]$.  Here
$\delta(t_1)=1$, if $t_1 - \Delta t \le t \le t_1$, where $t_1$ is the
time at which the burst started and  $\Delta t$ is its duration of the
burst, and zero  otherwise. The dotted and the  dashed lines represent
the luminosity functions when a burst of duration $\Delta t =1$~Gyr at
1 and 2~Gyr, respectively. As can  be seen, the effects of these bursts
are clearly visible in the slope of the luminosity function. On the
contrary, if the burst  of  star formation  is old enough, the
corresponding luminosity  functions  are barely  modified. This is
clearly borne out by inspecting  the solid lines  in Fig.~\ref{fig3},
which represent the luminosity functions for bursts occurring at 4, 6,
and 8~Gyr, respectively. It is evident that, since the old 
generations of stars are still producing new white dwarfs, a precise 
determination of the luminosity function at low luminsities could 
provide important constraints to the temporal evolution of the SFR.

\section{The axion case}
The most widely accepted solution to  the strong CP problem of quantum
chromodynamics consists  in the  introduction of  a new  symmetry, the
Peccei-Quinn  symmetry   \citep{pecc77a,pecc77b},  that  spontaneously
breaks at  some energy scale  and leads to  the formation of  a light
pseudo-scalar particle named axion \citep{wein78,wilc78}.  This energy
scale is not  fixed by the theory,  but it has to be  much larger than
the electroweak scale  to ensure a coupling between  axions and matter
weak enough to account  for the lack of detection up  to now. The mass
of axions and the energy scale are related by: $m_{\rm a} = 0.6(10^7\,
{\rm GeV}/f_{\rm  a})$~ eV.  Astrophysical and  cosmological arguments
limit  this mass  to the  range $10^{-6}  {\rm eV}  \la m_{\rm  a} \la
10^{-2}{\rm eV}$ (see \citet{turn90} and \citet{raff96} for a complete 
discussion of such bounds).

Axions couple to  photons, electrons and nucleons  with strengths that
depend on  the specific  implementation of the  Peccei-Quinn symmetry.
For instance,  in the KSVZ  -- or hadronic  model -- axions  couple to
hadrons and  photons only \citep{kim79,shif80},  while in the  DFSZ or
GUT  model   \citep{dine81,zhit80},  they   also  couple   to  charged
leptons. Stars can produce axions in their hot and dense interior in analogy
  as they release thermal  neutrinos \citep{raff96}.  These
particles can  freely escape from stars  and act as a  sink of energy.
In all the implementations, the degree of interaction between axions 
and matter/photons depends on the mass, a free parameter that is not fixed by the theory.
If  this  mass  is  large  enough, they  can  noticeably  modify  the
evolution of stars and reveal  their existence.  In particular, in the
case of white  dwarfs, axions can modify the cooling  rate. If the DFSZ 
implementation is adopted, the interaction of axions with electrons is dominant and 
is controlled by the axion fine constant, $\alpha = g^2_{\rm ae}/4\pi$, where  $g^2_{\rm ae}$ 
is the dimensionless Yukawa coupling and is related with the mass of the axion by
  $g^2_{\rm ae}= 2.8\times 10^{-4} m_{\rm a} \cos^2 \beta$, where $\tan \beta$ is 
the ratio between the two Higgs-field expectation values
and $m_{\rm a}$ is the mass of the axion in meV.

The first hint that axions could be acting in white dwarf stars  was provided 
by the secular drift of the period  of pulsation  of G117-B15A  \citep{iser92}, a  DA variable
white dwarf. The argument, see Fig. 1 of Isern et al. \emph{loc.cit.}, was that since DA variables are placed in the luminosity domain where the transition from a cooling dominated by neutrino bremsstrahlung, $\dot\epsilon _\nu \propto T^7$, to a cooling dominated 
by photons, $L_\nu \propto T^{2.5}$, axion bremsstrahlung, $\dot\epsilon _{\rm a} \propto T^4$, could provide the extra cooling term necessary to account for the secular drift observed in G117-B15A.

The second hint was provided by the white dwarf luminosity
function \citep{iser08a}.  As can be seen  in Fig.~\ref{fig1}, the
standard theoretical  luminosity functions predict a  slight excess of
stars  in  the  region  $M_{\rm  bol}   \sim  8  -  10$  that  can  be
satisfactorily removed  if DFSZ  axions with  $m_{\rm a}  \cos^2 \beta
\sim  2 -  10$~meV are considered.  These results, however, have to be regarded as
qualitative,  since the  uncertainties plaguing  the determination  of
both the observed and theoretical luminosity functions are still large
\citep{iser09,mill14}.

\section{Results and conclusions}

\begin{figure*}
\centering
\includegraphics[width=0.8\textwidth, clip=true, trim= 0cm 4cm 0cm 7cm]{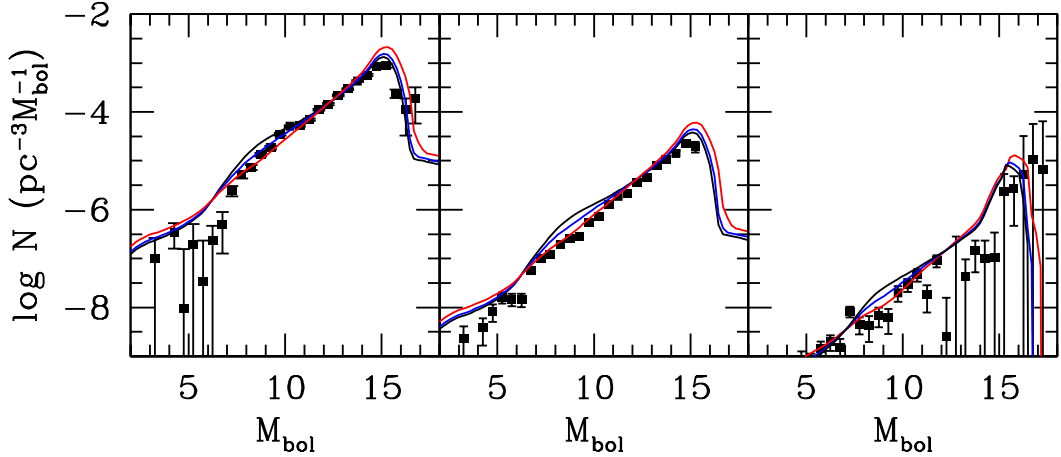}
\caption{From left  to right,  luminosity functions  of DA  and non--Da
  white dwarfs from  the SCSS catalogues for the thin  disk, the thick
  disk and  the halo  \citep{rowe11}.  The  solid lines  represent the
  luminosity  functions  obtained  disregarding  the  contribution  to
  cooling of  axions (black  lines) and  considering DFSZ  axions with
  coupling constants  $g_{\rm ae} =  2.24 \times 10^{-13}$ and  $ 4.48
  \times 10^{-13}$  (blue and red lines,  respectively), corresponding
  to masses $m_{\rm a} \cos^2 \beta \sim 8$ and $16$~meV.}
\label{fig4}
\end{figure*}

The luminosity function provided by \citet{rowe11} was obtained from a sample 
of the Super Cosmos Sky Survey (SCSS) containing $\sim$~10,000 brighter than $R \sim 19.75$ mag and with proper motions larger than $\mu \sim 0.05$ arc-sec yr$^{-1}$. This luminosity function is statistically significant in the range $6\, \lppr \, M_{\rm bol}\, \lppr\, 15$, but it does not provide  enough stars beyond the turnover, which is not an inconvenience since we are 
limiting ourselves to the interval $6\, \lppr \, M_{\rm bol}\, \lppr \, 12-13$. From a subsample of white dwarfs with tangential velocities $V_{\rm tan} > 200$~km yr$^{-1}$, they obtained a noticeably improved, by comparison with the previous ones, luminosity function of the halo. Finally, using statistical techniques, they were able to provide for the first time the white dwarf luminosity functions for the thin and thick disks.   Fig.~\ref{fig4} displays  these
luminosity functions.  The solid black line represents the theoretical
luminosity functions we have obtained disregarding axion emission and assuming
a constant SFR for  both the thin and the thick disks,  and a burst of short
duration for the halo.  In all  cases, an age of 13~Gyr has been
adopted (notice here that neither the adopted age nor the exact shape of the SFR are  relevant as far as we are only interested in the  bright branch of the luminosity  function).  As can
be seen, in all three cases  the standard model of cooling predicts an
excess of white dwarfs in the  same region around $M_{\rm bol} \simeq
10$, thus providing support to the hypothesis that the slight shortage
of stars in this  region is not a consequence of  the adopted SFR, but
of the existence of an additional cooling mechanism.


\begin{figure}
\centering
\includegraphics[width=\hsize,clip=true, trim= 0cm 1cm 0cm 3cm]{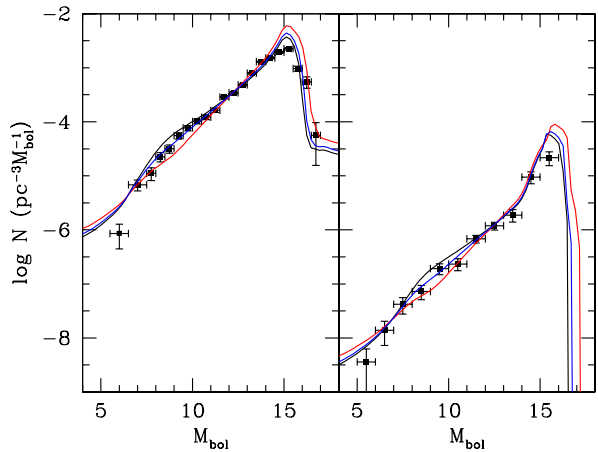}
\caption{From left  to right,  luminosity functions  of DA  and non-DA
  white dwarfs from  the SDSS Munn catalogue for the disk, thin and thick (left panel) and the halo (right panel). The meaning of the solid lines is the same as in Fig.~\ref{fig4}.}
\label{fig5}
\end{figure}


Recently, \citet{munn17} have improved the \citet{harr06} luminosity function of the disk using a sample that 
contains 2839 stars in the luminosity range $5.5\, \simeq \, M_{\rm bol}\, \simeq \, 17$ and tangential velocities  
in the range $40 \le v_{tan} \le 120$ km s$^{-1}$ to avoid contamination from subdwarfs. This luminosity function 
resolves the peak, since it contains 311 objects with bolometric magnitude $M_{\rm bol} \ge 15$ and improves the 
precision of the brighter bins. The scale height of white dwarfs was assumed to be constant and equal to 300 pc. 
They also constructed the luminosity function for the halo taking a sample of white dwarfs with $ 200 \le v_{\rm tan} 
\le 500$ km s$^{-1}$. Both luminosity functions are displayed in Fig.~\ref{fig5}. As in the previous case, visual 
inspection suggests a shortage of stars in the region $M_{\rm bol} \approx 10$ in both halo and disk functions. 

\begin{table}
	\centering
	\caption{Reduced $\chi^2$ obtained from the comparison between theoretical and observed white dwarf luminosity functions in the brightness interval $6\, \lappr \, M_{\rm bol} \, \lappr \, 12.5$ for different intensities of the coupling between electrons and axions.}
	\label{tab1}
	\begin{tabular}{lcccc} 
		\hline
		$g_{\rm ae}\times 10^{13}$   & 0.00 & 1.12 & 2.24 & 4.48 \\
                 $m_{\rm a} \cos \beta$ (meV) & 0  &     4   &   8    & 16    \\
		\hline
                   Rowell \& Hambly (2011)     &         &         &         &          \\
                   \hline
                   thin disk                             &18.59&15.33& 7.33 &21.52 \\
                   thick disk                           &30.72&26.61& 11.73& 1.43 \\
                   halo                                    &  3.31& 2.94 &  2.36 & 1.87  \\
                   \hline
                   Munn et al. (2017)                &          &         &         &          \\
                   \hline
                   thin+thick disk                  &  4.89 & 2.87 &  1.46 &  6.13 \\
                   halo                                    &  2.46 & 1.50 &  0.65 & 1.42 \\
		\hline
                   Kilic et al. (2017)                  &          &         &         &          \\
                   \hline
                    $\Phi_{200-900}$             &  3.25 & 2.09 &  1.05&  6.11  \\
                    $\Phi_{200-700}$             &  3.53 & 2.32 &  1.14&  5.80  \\
                    $\Phi_{200-500}$             &  4.02& 2.73  &  1.28&  2.26  \\
                    $\Phi_{300}$                     &  6.11& 4.95  &  2.26&   4.98 \\
		\hline
	\end{tabular}
\end{table}

 In Fig.~\ref{fig4}  we compare using  blue  and  red solid lines  the  luminosity functions 
of the thin and thick disk and halo obtained
including DFSZ axions with masses $m_{\rm  a} \cos^2 \beta \sim 8$ and
$16$~meV (equivalent to $g_{ae}\times 10^{13} =2.24$ and $4.48$)  
respectively,\footnote{The errors used in the calculation of $\chi^2$ in 
Table~\ref{tab1} are just the Poissonian errors. The systematic theoretical and observational 
uncertainties are larger and not well identified and therefore, they have been excluded. For this 
reason, we have always claimed that these luminosity functions provide a hint about the existence 
of axions, not a compelling evidence. The values of $\chi^2$ are used here only to illustrate 
a tendency.} with those obtained by \citet{rowe11}. As can be seen in Table~\ref{tab1}, the 
discrepancies  between  the  theoretical calculations and observations decrease in all three cases, 
if  axions  are  included.  In a similar way, Fig.~\ref{fig5} and Table~\ref{tab1} show that in the 
case of the luminosity functions obtained by \citet{munn17},
the inclusion of axions of moderate mass improves the concordance between the theory and observations in both, disk and halo.    

\begin{figure}
\centering
\includegraphics[width=\hsize]{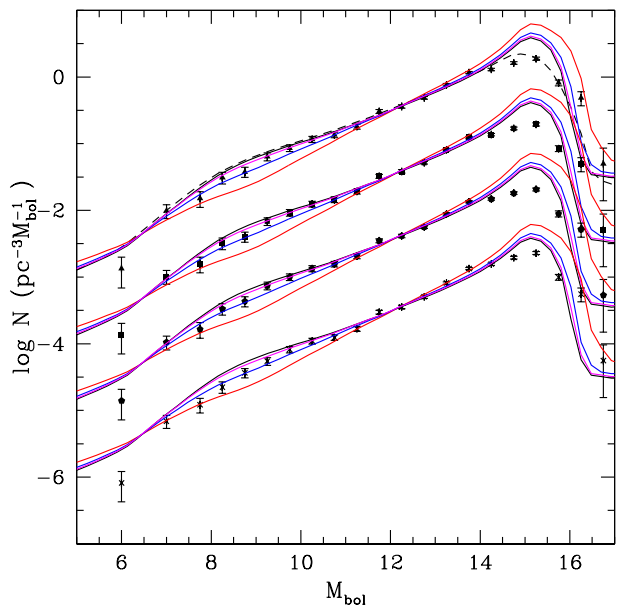}
\caption{White dwarf luminosity functions of the disk assuming different scale heights as proposed by \citet{kili17}. 
From top to bottom: $\Phi_{200-900}+3$ (triangles); $\Phi_{200-700}+2$ (squares);  $\Phi_{200-500}+1$ (pentagons);  
$\Phi_{300}$ (crosses). The  solid lines  represent the  luminosity  functions  obtained  assuming a constant SFR, 
no  axion cooling (black  lines) and DFSZ  axion cooling with coupling constants  
$g_{\rm ae} \times 10^{13} = 1.12, \, 2.24,\, 4.48 $  (magenta, blue and red lines,  respectively). 
The dashed line, on the top, represents a case with a constant SFR with an exponential tail and no axions. 
}
\label{fig6}
\end{figure}

As we have already mentioned, an important point is the influence of the assumed scale 
height on the bright branch of the luminosity function. \citet{kili17} constructed several 
observational  luminosity functions with assuming variable scale heights and concluded 
that the existing slight discrepancy in the region  $6\, \lappr \, M_{\rm bol} \, \lappr \, 12.5$ 
is caused by the use of a fixed scale height. In Fig.6 we have repeated the exercise of fitting 
their functions with our theoretical predictions with and without axions. Once more we have found, see Table~\ref{tab1}, that the inclusion of axions improves the agreement between theory and observations. All these luminosity functions have been obtained assuming a constant SFR and an age of the disk of 13 Gyr, which is obviously too large. However, as we have shown before, the luminosity function in the region we are considering is not strongly dependent on the age, provided it is large enough, nor on the shape of the old past SFR. As an example, in the top of Fig.6 we have plotted, with black dashed line, the luminosity function corresponding to 
$\Psi \propto (1+\exp [(t-t_0)/\tau])^{-1}, \tau = 3 \, {\rm Gyr},\, t_0= 9\,{\rm Gyr}$,
an SFR that roughly fits the peak of the function.

As can be deduced from Fig.~\ref{fig3}, the shape of the luminosity function could be
fitted just adding convenient star formation bursts. However, this not seems to be a plausible solution if the star formation history of the halo and discs are uncorrelated or just poorly correlated. Strictly speaking we cannot discard a conspiracy of Nature to produce similar features at the same place in different populations, but it seems more natural to invoke an intrinsic property of white dwarfs. Certainly the different cooling rates of DA and non-DA stars could be a solution but the results obtained in Fig.~\ref{fig2} together with the secular drift of the period of pulsation of ZZ Ceti stars mentioned in Section 2 make the axion hypothesis more attractive.
Obviously, these arguments do not  prove  the existence of axions but make them  (axions of DFSZ type with a mass in the range of few meV, 6 to 8)  a  good candidate  
to account  for such a behavior. This possibility  could be confirmed in a  near future when
the new  large cosmological  surveys (DES, LSST,\ldots),  together with the high precision  data from  the astrometric satellite  Gaia will provide  accurate and  reliable observational  white dwarf  luminosity functions.

Furthermore, if  DFSZ axions exist and  have masses in  the range of  a few
meV,  as suggested  by  this  study, they  will  introduce subtle  but
detectable changes  in the evolution  of stars as  is the case  of red
giants  \citep{viau13},  horizontal branch  stars  \citep{ayal14}, and globular clusters \citep{hans15,gold12,gold16}.
Additionally,  it  is expected  that  the  direct measurement  of  the
characteristic  cooling time  of  white dwarfs  employing the  secular
variation  of the  pulsation periods  of  ZZ~Ceti stars  will be  much
improved  in a  near future  \citep{cors12,cors12a,cors16}. This  will
also allow to obtain more precise estimates of this additional cooling
mechanism. Nevertheless, a definitive confirmation of the existence of
axions will  be provided  by the next  generation of  axioscopes, like
IAXO at CERN \citep{arme14}, that will  be able to directly detect the
emission of solar axions if their  mass, as suggested by the theory of
white dwarf cooling, is larger than 3~meV.
   
\section*{Acknowledgements}
This work  has been supported  by MINECO grants  ESP2015-66134-R (JI),
and AYA2014-59084-P (EG-B), by the  European Union FEDER funds, and by
grants 2014SGR1458 (JI)  and 2014SGR0038 (EG-B) and the CERCA program
 of  the Generalitat de Catalunya. One of us (JI) is very indebted to M. Lam and N. Rowell for the enlightening discussion about the influence of the scale height on the luminosity function.

\bibliographystyle{mnras} 
\bibliography{nou}

\begin{thebibliography}{}
\makeatletter
\relax
\def\mn@urlcharsother{\let\do\@makeother \do\$\do\&\do\#\do\^\do\_\do\%\do\~}
\def\mn@doi{\begingroup\mn@urlcharsother \@ifnextchar [ {\mn@doi@}
  {\mn@doi@[]}}
\def\mn@doi@[#1]#2{\def\@tempa{#1}\ifx\@tempa\@empty \href
  {http://dx.doi.org/#2} {doi:#2}\else \href {http://dx.doi.org/#2} {#1}\fi
  \endgroup}
\def\mn@eprint#1#2{\mn@eprint@#1:#2::\@nil}
\def\mn@eprint@arXiv#1{\href {http://arxiv.org/abs/#1} {{\tt arXiv:#1}}}
\def\mn@eprint@dblp#1{\href {http://dblp.uni-trier.de/rec/bibtex/#1.xml}
  {dblp:#1}}
\def\mn@eprint@#1:#2:#3:#4\@nil{\def\@tempa {#1}\def\@tempb {#2}\def\@tempc
  {#3}\ifx \@tempc \@empty \let \@tempc \@tempb \let \@tempb \@tempa \fi \ifx
  \@tempb \@empty \def\@tempb {arXiv}\fi \@ifundefined
  {mn@eprint@\@tempb}{\@tempb:\@tempc}{\expandafter \expandafter \csname
  mn@eprint@\@tempb\endcsname \expandafter{\@tempc}}}

\bibitem[\protect\citeauthoryear{{Althaus}, {C{\'o}rsico}, {Isern}  \&
  {Garc{\'{\i}}a-Berro}}{{Althaus} et~al.}{2010}]{alth10}
{Althaus} L.~G.,  {C{\'o}rsico} A.~H.,  {Isern} J.,   {Garc{\'{\i}}a-Berro} E.,
   2010, \mn@doi [\aapr] {10.1007/s00159-010-0033-1}, \href
  {http://cdsads.u-strasbg.fr/abs/2010A%26ARv..18..471A} {18, 471}

\bibitem[\protect\citeauthoryear{{Armengaud} et~al.,}{{Armengaud}
  et~al.}{2014}]{arme14}
{Armengaud} E.,  et~al., 2014, \mn@doi [Journal of Instrumentation]
  {10.1088/1748-0221/9/05/T05002}, \href
  {http://adsabs.harvard.edu/abs/2014JInst...9.5002A} {9, T05002}

\bibitem[\protect\citeauthoryear{{Ayala}, {Dom{\'{\i}}nguez}, {Giannotti},
  {Mirizzi}  \& {Straniero}}{{Ayala} et~al.}{2014}]{ayal14}
{Ayala} A.,  {Dom{\'{\i}}nguez} I.,  {Giannotti} M.,  {Mirizzi} A.,
  {Straniero} O.,  2014, \mn@doi [Physical Review Letters]
  {10.1103/PhysRevLett.113.191302}, \href
  {http://adsabs.harvard.edu/abs/2014PhRvL.113s1302A} {113, 191302}

\bibitem[\protect\citeauthoryear{{Barstow} et~al.,}{{Barstow}
  et~al.}{2014}]{bars14}
{Barstow} M.~A.,  et~al., 2014, preprint, \href
  {http://cdsads.u-strasbg.fr/abs/2014arXiv1407.6163B} {} (\mn@eprint {arXiv}
  {1407.6163})

\bibitem[\protect\citeauthoryear{{Catal{\'a}n}, {Isern}, {Garc{\'{\i}}a-Berro}
  \& {Ribas}}{{Catal{\'a}n} et~al.}{2008}]{cata08}
{Catal{\'a}n} S.,  {Isern} J.,  {Garc{\'{\i}}a-Berro} E.,   {Ribas} I.,  2008,
  \mn@doi [\mnras] {10.1111/j.1365-2966.2008.13356.x}, \href
  {http://cdsads.u-strasbg.fr/abs/2008MNRAS.387.1693C} {387, 1693}

\bibitem[\protect\citeauthoryear{{Cignoni}, {Degl'Innocenti}, {Prada Moroni}
  \& {Shore}}{{Cignoni} et~al.}{2006}]{cign06}
{Cignoni} M.,  {Degl'Innocenti} S.,  {Prada Moroni} P.~G.,   {Shore} S.~N.,
  2006, \mn@doi [\aap] {10.1051/0004-6361:20065645}, \href
  {http://cdsads.u-strasbg.fr/abs/2006A%26A...459..783C} {459, 783}

\bibitem[\protect\citeauthoryear{{C{\'o}rsico}, {Althaus}, {Romero}, {Mukadam},
  {Garc{\'{\i}}a-Berro}, {Isern}, {Kepler}  \& {Corti}}{{C{\'o}rsico}
  et~al.}{2012a}]{cors12}
{C{\'o}rsico} A.~H.,  {Althaus} L.~G.,  {Romero} A.~D.,  {Mukadam} A.~S.,
  {Garc{\'{\i}}a-Berro} E.,  {Isern} J.,  {Kepler} S.~O.,   {Corti} M.~A.,
  2012a, \mn@doi [\jcap] {10.1088/1475-7516/2012/12/010}, \href
  {http://adsabs.harvard.edu/abs/2012JCAP...12..010C} {12, 010}

\bibitem[\protect\citeauthoryear{{C{\'o}rsico}, {Althaus}, {Miller Bertolami},
  {Romero}, {Garc{\'{\i}}a-Berro}, {Isern}  \& {Kepler}}{{C{\'o}rsico}
  et~al.}{2012b}]{cors12a}
{C{\'o}rsico} A.~H.,  {Althaus} L.~G.,  {Miller Bertolami} M.~M.,  {Romero}
  A.~D.,  {Garc{\'{\i}}a-Berro} E.,  {Isern} J.,   {Kepler} S.~O.,  2012b,
  \mn@doi [\mnras] {10.1111/j.1365-2966.2012.21401.x}, \href
  {http://adsabs.harvard.edu/abs/2012MNRAS.424.2792C} {424, 2792}

\bibitem[\protect\citeauthoryear{{C{\'o}rsico} et~al.,}{{C{\'o}rsico}
  et~al.}{2016}]{cors16}
{C{\'o}rsico} A.~H.,  et~al., 2016, \mn@doi [\jcap]
  {10.1088/1475-7516/2016/07/036}, \href
  {http://adsabs.harvard.edu/abs/2016JCAP...07..036C} {7, 036}

\bibitem[\protect\citeauthoryear{{DeGennaro}, {von Hippel}, {Winget}, {Kepler},
  {Nitta}, {Koester}  \& {Althaus}}{{DeGennaro} et~al.}{2008}]{dege08}
{DeGennaro} S.,  {von Hippel} T.,  {Winget} D.~E.,  {Kepler} S.~O.,  {Nitta}
  A.,  {Koester} D.,   {Althaus} L.,  2008, \mn@doi [\aj]
  {10.1088/0004-6256/135/1/1}, \href
  {http://adsabs.harvard.edu/abs/2008AJ....135....1D} {135, 1}

\bibitem[\protect\citeauthoryear{{Dine}, {Fischler}  \& {Srednicki}}{{Dine}
  et~al.}{1981}]{dine81}
{Dine} M.,  {Fischler} W.,   {Srednicki} M.,  1981, \mn@doi [Physics Letters B]
  {10.1016/0370-2693(81)90590-6}, \href
  {http://cdsads.u-strasbg.fr/abs/1981PhLB..104..199D} {104, 199}

\bibitem[\protect\citeauthoryear{{Evans}}{{Evans}}{1992}]{evan92}
{Evans} D.~W.,  1992, \mn@doi [\mnras] {10.1093/mnras/255.3.521}, \href
  {http://adsabs.harvard.edu/abs/1992MNRAS.255..521E} {255, 521}

\bibitem[\protect\citeauthoryear{{Fontaine}, {Brassard}  \&
  {Bergeron}}{{Fontaine} et~al.}{2001}]{font01}
{Fontaine} G.,  {Brassard} P.,   {Bergeron} P.,  2001, \mn@doi [\pasp]
  {10.1086/319535}, \href {http://cdsads.u-strasbg.fr/abs/2001PASP..113..409F}
  {113, 409}

\bibitem[\protect\citeauthoryear{{Garc{\'{\i}}a-Berro} \&
  {Oswalt}}{{Garc{\'{\i}}a-Berro} \& {Oswalt}}{2016}]{garc16}
{Garc{\'{\i}}a-Berro} E.,  {Oswalt} T.~D.,  2016, \mn@doi [\nar]
  {10.1016/j.newar.2016.08.001}, \href
  {http://adsabs.harvard.edu/abs/2016NewAR..72....1G} {72, 1}

\bibitem[\protect\citeauthoryear{{Garcia-Berro}, {Hernanz}, {Isern}  \&
  {Mochkovitch}}{{Garcia-Berro} et~al.}{1995}]{garc95}
{Garcia-Berro} E.,  {Hernanz} M.,  {Isern} J.,   {Mochkovitch} R.,  1995,
  \mn@doi [\mnras] {10.1093/mnras/277.3.801}, \href
  {http://adsabs.harvard.edu/abs/1995MNRAS.277..801G} {277, 801}

\bibitem[\protect\citeauthoryear{{Garc{\'{\i}}a-Berro}, {Torres}, {Althaus}  \&
  {Miller Bertolami}}{{Garc{\'{\i}}a-Berro} et~al.}{2014}]{garc14}
{Garc{\'{\i}}a-Berro} E.,  {Torres} S.,  {Althaus} L.~G.,   {Miller Bertolami}
  M.~M.,  2014, \mn@doi [\aap] {10.1051/0004-6361/201424652}, \href
  {http://adsabs.harvard.edu/abs/2014A%26A...571A..56G} {571, A56}

\bibitem[\protect\citeauthoryear{{Goldsbury} et~al.,}{{Goldsbury}
  et~al.}{2012}]{gold12}
{Goldsbury} R.,  et~al., 2012, \mn@doi [\apj] {10.1088/0004-637X/760/1/78},
  \href {http://adsabs.harvard.edu/abs/2012ApJ...760...78G} {760, 78}

\bibitem[\protect\citeauthoryear{{Goldsbury}, {Heyl}, {Richer}, {Kalirai}  \&
  {Tremblay}}{{Goldsbury} et~al.}{2016}]{gold16}
{Goldsbury} R.,  {Heyl} J.,  {Richer} H.~B.,  {Kalirai} J.~S.,   {Tremblay}
  P.~E.,  2016, \mn@doi [\apj] {10.3847/0004-637X/821/1/27}, \href
  {http://adsabs.harvard.edu/abs/2016ApJ...821...27G} {821, 27}

\bibitem[\protect\citeauthoryear{{Gratton}, {Carretta}  \&
  {Bragaglia}}{{Gratton} et~al.}{2012}]{grat12}
{Gratton} R.~G.,  {Carretta} E.,   {Bragaglia} A.,  2012, \mn@doi [\aapr]
  {10.1007/s00159-012-0050-3}, \href
  {http://adsabs.harvard.edu/abs/2012A%26ARv..20...50G} {20, 50}

\bibitem[\protect\citeauthoryear{{Hansen}}{{Hansen}}{2004}]{hans04}
{Hansen} B.,  2004, \mn@doi [\physrep] {10.1016/j.physrep.2004.07.001}, \href
  {http://adsabs.harvard.edu/abs/2004PhR...399....1H} {399, 1}

\bibitem[\protect\citeauthoryear{{Hansen} \& {Liebert}}{{Hansen} \&
  {Liebert}}{2003}]{hans03}
{Hansen} B.~M.~S.,  {Liebert} J.,  2003, \mn@doi [\araa]
  {10.1146/annurev.astro.41.081401.155117}, \href
  {http://cdsads.u-strasbg.fr/abs/2003ARA%26A..41..465H} {41, 465}

\bibitem[\protect\citeauthoryear{{Hansen}, {Richer}, {Kalirai}, {Goldsbury},
  {Frewen}  \& {Heyl}}{{Hansen} et~al.}{2015}]{hans15}
{Hansen} B.~M.~S.,  {Richer} H.,  {Kalirai} J.,  {Goldsbury} R.,  {Frewen} S.,
   {Heyl} J.,  2015, \mn@doi [\apj] {10.1088/0004-637X/809/2/141}, \href
  {http://adsabs.harvard.edu/abs/2015ApJ...809..141H} {809, 141}

\bibitem[\protect\citeauthoryear{{Harris} et~al.,}{{Harris}
  et~al.}{2006}]{harr06}
{Harris} H.~C.,  et~al., 2006, \mn@doi [\aj] {10.1086/497966}, \href
  {http://adsabs.harvard.edu/abs/2006AJ....131..571H} {131, 571}

\bibitem[\protect\citeauthoryear{{Isern} \& {Garc{\'{\i}}a-Berro}}{{Isern} \&
  {Garc{\'{\i}}a-Berro}}{2008}]{iser08}
{Isern} J.,  {Garc{\'{\i}}a-Berro} E.,  2008, \memsai, \href
  {http://cdsads.u-strasbg.fr/abs/2008MmSAI..79..545I} {79, 545}

\bibitem[\protect\citeauthoryear{{Isern}, {Hernanz}  \& {Garcia-Berro}}{{Isern}
  et~al.}{1992}]{iser92}
{Isern} J.,  {Hernanz} M.,   {Garcia-Berro} E.,  1992, \mn@doi [\apjl]
  {10.1086/186416}, \href {http://adsabs.harvard.edu/abs/1992ApJ...392L..23I}
  {392, L23}

\bibitem[\protect\citeauthoryear{{Isern}, {Garc{\'{\i}}a-Berro}, {Hernanz}  \&
  {Mochkovitch}}{{Isern} et~al.}{1998a}]{iser98a}
{Isern} J.,  {Garc{\'{\i}}a-Berro} E.,  {Hernanz} M.,   {Mochkovitch} R.,
  1998a, \mn@doi [Journal of Physics Condensed Matter]
  {10.1088/0953-8984/10/49/015}, \href
  {http://cdsads.u-strasbg.fr/abs/1998JPCM...1011263I} {10, 11263}

\bibitem[\protect\citeauthoryear{{Isern}, {Garc{\'{\i}}a-Berro}, {Hernanz},
  {Mochkovitch}  \& {Torres}}{{Isern} et~al.}{1998b}]{iser98}
{Isern} J.,  {Garc{\'{\i}}a-Berro} E.,  {Hernanz} M.,  {Mochkovitch} R.,
  {Torres} S.,  1998b, \mn@doi [\apj] {10.1086/305977}, \href
  {http://adsabs.harvard.edu/abs/1998ApJ...503..239I} {503, 239}

\bibitem[\protect\citeauthoryear{{Isern}, {Garc{\'{\i}}a-Berro}, {Torres}  \&
  {Catal{\'a}n}}{{Isern} et~al.}{2008}]{iser08a}
{Isern} J.,  {Garc{\'{\i}}a-Berro} E.,  {Torres} S.,   {Catal{\'a}n} S.,  2008,
  \mn@doi [\apjl] {10.1086/591042}, \href
  {http://adsabs.harvard.edu/abs/2008ApJ...682L.109I} {682, L109}

\bibitem[\protect\citeauthoryear{{Isern}, {Catal{\'a}n}, {Garc{\'{\i}}a-Berro}
  \& {Torres}}{{Isern} et~al.}{2009}]{iser09}
{Isern} J.,  {Catal{\'a}n} S.,  {Garc{\'{\i}}a-Berro} E.,   {Torres} S.,  2009,
  \mn@doi [Journal of Physics Conference Series]
  {10.1088/1742-6596/172/1/012005}, \href
  {http://cdsads.u-strasbg.fr/abs/2009JPhCS.172a2005I} {172, 012005}

\bibitem[\protect\citeauthoryear{{Isern}, {Artigas}  \&
  {Garc{\'{\i}}a-Berro}}{{Isern} et~al.}{2013}]{iser13}
{Isern} J.,  {Artigas} A.,   {Garc{\'{\i}}a-Berro} E.,  2013, in European
  Physical Journal Web of Conferences. p. 05002 (\mn@eprint {arXiv}
  {1212.0806}), \mn@doi{10.1051/epjconf/20134305002}

\bibitem[\protect\citeauthoryear{{Kilic}, {Munn}, {Harris}, {von Hippel},
  {Liebert}, {Williams}, {Jeffery}  \& {DeGennaro}}{{Kilic}
  et~al.}{2017}]{kili17}
{Kilic} M.,  {Munn} J.~A.,  {Harris} H.~C.,  {von Hippel} T.,  {Liebert} J.~W.,
   {Williams} K.~A.,  {Jeffery} E.,   {DeGennaro} S.,  2017, \mn@doi [\apj]
  {10.3847/1538-4357/aa62a5}, \href
  {http://adsabs.harvard.edu/abs/2017ApJ...837..162K} {837, 162}

\bibitem[\protect\citeauthoryear{{Kim}}{{Kim}}{1979}]{kim79}
{Kim} J.~E.,  1979, \mn@doi [Physical Review Letters]
  {10.1103/PhysRevLett.43.103}, \href
  {http://cdsads.u-strasbg.fr/abs/1979PhRvL..43..103K} {43, 103}

\bibitem[\protect\citeauthoryear{{K{\i}z{\i}ltan}, {Baumgardt}  \&
  {Loeb}}{{K{\i}z{\i}ltan} et~al.}{2017}]{kizi17}
{K{\i}z{\i}ltan} B.,  {Baumgardt} H.,   {Loeb} A.,  2017, \mn@doi [\nat]
  {10.1038/nature21361}, \href
  {http://adsabs.harvard.edu/abs/2017Natur.542..203K} {542, 203}

\bibitem[\protect\citeauthoryear{{Knox}, {Hawkins}  \& {Hambly}}{{Knox}
  et~al.}{1999}]{knox99}
{Knox} R.~A.,  {Hawkins} M.~R.~S.,   {Hambly} N.~C.,  1999, \mn@doi [\mnras]
  {10.1046/j.1365-8711.1999.02625.x}, \href
  {http://adsabs.harvard.edu/abs/1999MNRAS.306..736K} {306, 736}

\bibitem[\protect\citeauthoryear{{Koester}}{{Koester}}{2002}]{koes02}
{Koester} D.,  2002, \mn@doi [\aapr] {10.1007/s001590100015}, \href
  {http://cdsads.u-strasbg.fr/abs/2002A%26ARv..11...33K} {11, 33}

\bibitem[\protect\citeauthoryear{{Koester} \& {Chanmugam}}{{Koester} \&
  {Chanmugam}}{1990}]{koes90}
{Koester} D.,  {Chanmugam} G.,  1990, \mn@doi [Reports on Progress in Physics]
  {10.1088/0034-4885/53/7/001}, \href
  {http://cdsads.u-strasbg.fr/abs/1990RPPh...53..837K} {53, 837}

\bibitem[\protect\citeauthoryear{{Krzesinski}, {Kleinman}, {Nitta},
  {H{\"u}gelmeyer}, {Dreizler}, {Liebert}  \& {Harris}}{{Krzesinski}
  et~al.}{2009}]{krze09}
{Krzesinski} J.,  {Kleinman} S.~J.,  {Nitta} A.,  {H{\"u}gelmeyer} S.,
  {Dreizler} S.,  {Liebert} J.,   {Harris} H.,  2009, \mn@doi [\aap]
  {10.1051/0004-6361/200912094}, \href
  {http://cdsads.u-strasbg.fr/abs/2009A%26A...508..339K} {508, 339}

\bibitem[\protect\citeauthoryear{{Leggett}, {Ruiz}  \& {Bergeron}}{{Leggett}
  et~al.}{1998}]{legg98}
{Leggett} S.~K.,  {Ruiz} M.~T.,   {Bergeron} P.,  1998, \mn@doi [\apj]
  {10.1086/305463}, \href {http://adsabs.harvard.edu/abs/1998ApJ...497..294L}
  {497, 294}

\bibitem[\protect\citeauthoryear{{Liebert}, {Dahn}  \& {Monet}}{{Liebert}
  et~al.}{1988}]{lieb88}
{Liebert} J.,  {Dahn} C.~C.,   {Monet} D.~G.,  1988, \mn@doi [\apj]
  {10.1086/166699}, \href {http://cdsads.u-strasbg.fr/abs/1988ApJ...332..891L}
  {332, 891}

\bibitem[\protect\citeauthoryear{{Limoges} \& {Bergeron}}{{Limoges} \&
  {Bergeron}}{2010}]{limo10}
{Limoges} M.-M.,  {Bergeron} P.,  2010, \mn@doi [\apj]
  {10.1088/0004-637X/714/2/1037}, \href
  {http://adsabs.harvard.edu/abs/2010ApJ...714.1037L} {714, 1037}

\bibitem[\protect\citeauthoryear{{Miller Bertolami}, {Melendez}, {Althaus}  \&
  {Isern}}{{Miller Bertolami} et~al.}{2014}]{mill14}
{Miller Bertolami} M.~M.,  {Melendez} B.~E.,  {Althaus} L.~G.,   {Isern} J.,
  2014, \mn@doi [\jcap] {10.1088/1475-7516/2014/10/069}, \href
  {http://cdsads.u-strasbg.fr/abs/2014JCAP...10..069M} {10, 069}

\bibitem[\protect\citeauthoryear{{Munn} et~al.,}{{Munn} et~al.}{2017}]{munn17}
{Munn} J.~A.,  et~al., 2017, \mn@doi [\aj] {10.3847/1538-3881/153/1/10}, \href
  {http://adsabs.harvard.edu/abs/2017AJ....153...10M} {153, 10}

\bibitem[\protect\citeauthoryear{{Noh} \& {Scalo}}{{Noh} \&
  {Scalo}}{1990}]{noh90}
{Noh} H.-R.,  {Scalo} J.,  1990, \mn@doi [\apj] {10.1086/168562}, \href
  {http://adsabs.harvard.edu/abs/1990ApJ...352..605N} {352, 605}

\bibitem[\protect\citeauthoryear{{Oswalt}, {Smith}, {Wood}  \&
  {Hintzen}}{{Oswalt} et~al.}{1996}]{oswa96}
{Oswalt} T.~D.,  {Smith} J.~A.,  {Wood} M.~A.,   {Hintzen} P.,  1996, \mn@doi
  [\nat] {10.1038/382692a0}, \href
  {http://adsabs.harvard.edu/abs/1996Natur.382..692O} {382, 692}

\bibitem[\protect\citeauthoryear{{Peccei} \& {Quinn}}{{Peccei} \&
  {Quinn}}{1977a}]{pecc77b}
{Peccei} R.~D.,  {Quinn} H.~R.,  1977a, \mn@doi [\prd]
  {10.1103/PhysRevD.16.1791}, \href
  {http://cdsads.u-strasbg.fr/abs/1977PhRvD..16.1791P} {16, 1791}

\bibitem[\protect\citeauthoryear{{Peccei} \& {Quinn}}{{Peccei} \&
  {Quinn}}{1977b}]{pecc77a}
{Peccei} R.~D.,  {Quinn} H.~R.,  1977b, \mn@doi [Physical Review Letters]
  {10.1103/PhysRevLett.38.1440}, \href
  {http://cdsads.u-strasbg.fr/abs/1977PhRvL..38.1440P} {38, 1440}

\bibitem[\protect\citeauthoryear{{Raffelt}}{{Raffelt}}{1996}]{raff96}
{Raffelt} G.~G.,  1996, {Stars as laboratories for fundamental physics : the
  astrophysics of neutrinos, axions, and other weakly interacting particles}.
University of Chicago Press

\bibitem[\protect\citeauthoryear{{Reid}}{{Reid}}{2005}]{reid05}
{Reid} I.~N.,  2005, \mn@doi [\araa] {10.1146/annurev.astro.43.072103.150623},
  \href {http://adsabs.harvard.edu/abs/2005ARA%26A..43..247R} {43, 247}

\bibitem[\protect\citeauthoryear{{Renedo}, {Althaus}, {Miller Bertolami},
  {Romero}, {C{\'o}rsico}, {Rohrmann}  \& {Garc{\'{\i}}a-Berro}}{{Renedo}
  et~al.}{2010}]{rene10}
{Renedo} I.,  {Althaus} L.~G.,  {Miller Bertolami} M.~M.,  {Romero} A.~D.,
  {C{\'o}rsico} A.~H.,  {Rohrmann} R.~D.,   {Garc{\'{\i}}a-Berro} E.,  2010,
  \mn@doi [\apj] {10.1088/0004-637X/717/1/183}, \href
  {http://adsabs.harvard.edu/abs/2010ApJ...717..183R} {717, 183}

\bibitem[\protect\citeauthoryear{{Rowell}}{{Rowell}}{2013}]{rowe13}
{Rowell} N.,  2013, \mn@doi [\mnras] {10.1093/mnras/stt1110}, \href
  {http://cdsads.u-strasbg.fr/abs/2013MNRAS.434.1549R} {434, 1549}

\bibitem[\protect\citeauthoryear{{Rowell} \& {Hambly}}{{Rowell} \&
  {Hambly}}{2011}]{rowe11}
{Rowell} N.,  {Hambly} N.~C.,  2011, \mn@doi [\mnras]
  {10.1111/j.1365-2966.2011.18976.x}, \href
  {http://cdsads.u-strasbg.fr/abs/2011MNRAS.417...93R} {417, 93}

\bibitem[\protect\citeauthoryear{{Salaris}, {Cassisi}, {Pietrinferni},
  {Kowalski}  \& {Isern}}{{Salaris} et~al.}{2010}]{sala10}
{Salaris} M.,  {Cassisi} S.,  {Pietrinferni} A.,  {Kowalski} P.~M.,   {Isern}
  J.,  2010, \mn@doi [\apj] {10.1088/0004-637X/716/2/1241}, \href
  {http://adsabs.harvard.edu/abs/2010ApJ...716.1241S} {716, 1241}

\bibitem[\protect\citeauthoryear{{Shifman}, {Vainshtein}  \&
  {Zakharov}}{{Shifman} et~al.}{1980}]{shif80}
{Shifman} M.~A.,  {Vainshtein} A.~I.,   {Zakharov} V.~I.,  1980, \mn@doi
  [Nuclear Physics B] {10.1016/0550-3213(80)90389-2}, \href
  {http://cdsads.u-strasbg.fr/abs/1980NuPhB.163...46S} {163, 46}

\bibitem[\protect\citeauthoryear{{Turner}}{{Turner}}{1990}]{turn90}
{Turner} M.~S.,  1990, \mn@doi [\physrep] {10.1016/0370-1573(90)90172-X}, \href
  {http://adsabs.harvard.edu/abs/1990PhR...197...67T} {197, 67}

\bibitem[\protect\citeauthoryear{{Viaux}, {Catelan}, {Stetson}, {Raffelt},
  {Redondo}, {Valcarce}  \& {Weiss}}{{Viaux} et~al.}{2013}]{viau13}
{Viaux} N.,  {Catelan} M.,  {Stetson} P.~B.,  {Raffelt} G.~G.,  {Redondo} J.,
  {Valcarce} A.~A.~R.,   {Weiss} A.,  2013, \mn@doi [Physical Review Letters]
  {10.1103/PhysRevLett.111.231301}, \href
  {http://adsabs.harvard.edu/abs/2013PhRvL.111w1301V} {111, 231301}

\bibitem[\protect\citeauthoryear{{Weidemann}}{{Weidemann}}{1968}]{weid68}
{Weidemann} V.,  1968, \mn@doi [\araa] {10.1146/annurev.aa.06.090168.002031},
  \href {http://adsabs.harvard.edu/abs/1968ARA%26A...6..351W} {6, 351}

\bibitem[\protect\citeauthoryear{{Weinberg}}{{Weinberg}}{1978}]{wein78}
{Weinberg} S.,  1978, \mn@doi [Physical Review Letters]
  {10.1103/PhysRevLett.40.223}, \href
  {http://cdsads.u-strasbg.fr/abs/1978PhRvL..40..223W} {40, 223}

\bibitem[\protect\citeauthoryear{{Wilczek}}{{Wilczek}}{1978}]{wilc78}
{Wilczek} F.,  1978, \mn@doi [Physical Review Letters]
  {10.1103/PhysRevLett.40.279}, \href
  {http://cdsads.u-strasbg.fr/abs/1978PhRvL..40..279W} {40, 279}

\bibitem[\protect\citeauthoryear{{Zhitnitskii}}{{Zhitnitskii}}{1980}]{zhit80}
{Zhitnitskii} A.~P.,  1980, \mn@doi [Soviet Journal Nuclear Physics]
  {10.1103/}, \href {http://cdsads.u-strasbg.fr/abs/1978PhRvL..40..279W} {31,
  260}

\makeatother
\end{thebibliography}

\end{document}